\documentclass[nofootinbib,prd,twocolumn,showpacs,showkeys,preprintnumbers]{revtex4-1}
\usepackage{hyperref,amssymb,amsmath,mathrsfs,bm,graphicx}
\begin{document}
\title { New definition of complexity for  self--gravitating  fluid distributions: The   spherically symmetric, static case.}
\author{L. Herrera}
\email{lherrera@usal.es}
\affiliation{Instituto Universitario de F\'isica
Fundamental y Matem\'aticas, Universidad de Salamanca, Salamanca 37007, Spain}
\date{\today}
\begin{abstract}
We put forward  a new definition of complexity, for static and spherically symmetric self--gravitating systems,  based on  a quantity, hereafter referred to as complexity factor, that    appears in  the orthogonal splitting of the Riemann tensor, in the context of general relativity. 
We start by assuming that the homogeneous (in the energy density) fluid, with isotropic pressure  is endowed with minimal complexity. For this kind of fluid distribution, the value of complexity factor is zero. So, the rationale behind our proposal  for the definition of complexity factor stems from the fact that it measures the departure, in the value of the active gravitational mass (Tolman mass), with respect to its value for  a zero complexity system. Such departure is   produced by a specific combination of energy density inhomogeneity and pressure anisotropy. Thus, zero  complexity factor may also be found in self--gravitating systems with inhomogeneous energy density and anisotropic pressure, provided the effects of these two factors, on the complexity factor, cancel each other.  Some   exact interior solutions to the Einstein equations satisfying the zero complexity criterium are found, and prospective applications of this newly defined concept, to the study of the structure and evolution of compact objects, are discussed.
 \end{abstract}
\date{\today}
\maketitle

\section{Introduction}

Many efforts have been devoted in the past  towards
a rigorous definition of complexity in different branches of science, although there is
not yet a consensus on a precise definition (see \cite{0, 1, 1bis, 2, 2bis, 2bisbis, 3, 3bis, 3bisbis, 4, 4bis, 6bis} and references therein). 

Among  the many definitions that have been proposed so far, most of them resort to concepts
such as information and entropy, and are based on the intuitive idea that  complexity should, somehow, measure a basic property describing  the  structures existing within a  system (not necessarily a physical one).

Thus, when dealing with a situation that intuitively is judged as ``complex'', we need to be able 
 to quantify this complexity, by defining
an observable measuring it. It is the purpose of this work to  define one such quantity, for self--gravitating systems in the context of general relativity.

Usually, the notion of complexity in physics 
starts by considering the perfect crystal  (periodic behaviour) and the isolated
ideal gas (random behaviour), as examples of simplest models and therefore
as systems with zero complexity. 

A perfect crystal is completely ordered and the
atoms are arranged following specific rules of symmetry.
The probability distribution for the states accessible
to the perfect crystal is centered around a
prevailing state of perfect symmetry, i.e.  it has low information content.

 On the other hand,
the isolated ideal gas is completely disordered. The
system can be found in any of its accessible states
with the same probability, i.e.  it has  a maximum information. 

Therefore, since these
two simple systems are extreme in the scale of “order”
and “information”, it is evident that the definition
of complexity, must include another factors, besides  
order or information.

 In \cite{3}  the concept of ``disequilibrium'' was introduced, which measures the ``distance'' from the equiprobable distribution of the accessible states of the system. Thus,  it would be different from zero
if there are privileged, or more probable, states among
those accessible. Therefore, ``disequilibrium'' would be
maximum for  a perfect crystal, since it is   far from an equidistribution
among the accessible states, whereas it would be zero for the ideal gas.

A compromise between these two concepts ( ``disequilibrium''   and information) is reached by defining the complexity through a quantity, which is  a product of  these two concepts (see \cite{3} for details). Doing so, one ensures that complexity vanishes for, both, the perfect crystal and the ideal gas.

It should be reminded that a definition of complexity, based on the work developed by Lopez--Ruiz and collaborators \cite{3, 4},  has already been proposed  for self--gravitating systems \cite{5, 6, 7, 8, 9, 10}. However, as we shall see below, the definition we propose here is quite different.

Indeed, our definition, although intuitively associated to the very concept of ``structure'' within the fluid distribution,  is not related (at least not directly) to information or disequilibrium, rather it stems from the basic assumption that the simplest system (at least one of them) is represented by the homogeneous fluid with isotropic pressure. Having assumed this fact as a natural definition of a vanishing complexity system, the very definition of complexity will emerge in  the development of  the fundamental theory of self--gravitating compact objects, in the context of general relativity.

The basic motivation for our enedeavour, resides in the fact that, the  definition of complexity for self--gravitating systems  proposed in \cite{5, 6, 7, 8, 9, 10}, contains two features  which, in our opinion, are unsatisfactory.

 On the one hand, the probability distribution   which appear in the definition of  ``disequilibrium''   and information, is replaced, in \cite{5, 6, 7, 8, 9, 10}, by the energy density of the fluid distribution.  This has been  justified by the statement that the energy density is related to the probability of finding some particles at a given defined location inside the star, or, plainly, by the fact that  it proved difficult to suggest a better alternative from the available physical quantities. 

On the other hand, in the above mentioned definition of complexity, only intervenes the energy density of the fluid, whereas other important variables such as the pressure components of the energy--momentum tensor, which we expect to play an important role in the formation of any structure within the fluid distribution,  are absent. This is a particularly unsatisfactory situation, if we recall that the very concept of complexity should, somehow, be associated to generic properties of the structure of the fluid.

For the reasons expressed above, we intend in this work to introduce a new concept of complexity, for self--gravitating systems. The variable responsible for measuring complexity, which we call the complexity factor, appears in the orthogonal splitting of the Riemann tensor, and the justification for such a proposition, roughly speaking, is as follows (see section IV for a more detailed discussion).

For a static fluid distribution, the simplest system is represented by a homogeneous (in the energy density), locally  isotropic fluid (principal stresses equal). So we assign a zero value of  the complexity factor for such a distribution. Next, we recall the concept of the Tolman mass \cite{11}, which may be interpreted as the ``active''  gravitational mass, and as we shall see may be expressed, for an arbitrary distribution,  through its value for the zero complexity case plus two terms depending on the energy density inhomogeneity and pressure anisotropy, respectively.  These latter terms, in its turn may be expressed through a single scalar function that we call the complexity factor. It obviously vanishes when the fluid is homogeneous in the energy density,  and isotropic in pressure, but also may vanish when the two terms  containing density inhomogeneity and anisotropic pressure, cancel each other. Thus as in \cite{3}, vanishing complexity  may correspond to very different systems.

The manuscript is organized as follows: In the next section we introduce all the variables and  conventions used throughout the paper. In section III we briefly review the orthogonal splitting of the Riemann tensor and the origin of the so called structure scalars, one of which would play the role of the complexity factor. After introducing the  complexity factor  in section IV, we shall display  in section V, different exact solutions  to the Einstein equations with  vanishing complexity factor. Finally, a summary of the obtained results  as well as prospective applications and extensions of the complexity factor, to more general situations, are presented in section VI. 

\section{THE VARIABLES AND THE EQUATIONS}
 
In this section we shall present the physical variables and the relevant equations for describing a static self--gravitating locally anisotropic fluid. For more details see \cite{13, 14, 16}.
\subsection{Einstein equations}

We consider spherically symmetric distributions of  static 
fluid, which for the sake of completeness we assume to be locally anisotropic and  bounded by a
spherical surface $\Sigma$.

\noindent
The line element is given in Schwarzschild--like  coordinates by

\begin{equation}
ds^2=e^{\nu} dt^2 - e^{\lambda} dr^2 -
r^2 \left( d\theta^2 + \sin^2\theta d\phi^2 \right),
\label{metric}
\end{equation}

\noindent
where $\nu(r)$ and $\lambda(r)$ are functions of their arguments. We
number the coordinates: $x^0=t; \, x^1=r; \, x^2=\theta; \, x^3=\phi$.

\noindent
The metric (\ref{metric}) has to satisfy Einstein field equations

\begin{equation}
G^\nu_\mu=8\pi T^\nu_\mu,
\label{Efeq}
\end{equation}

\noindent

In order to give physical significance to the $T^{\mu}_{\nu}$ components
we apply the Bondi approach \cite{17}.

\noindent
Thus, following Bondi, let us introduce purely locally Minkowski
coordinates ($\tau, x, y, z$)

$$d\tau=e^{\nu/2}dt\,;\quad\,dx=e^{\lambda/2}dr\,;\quad\,
dy=rd\theta\,;\quad\, dz=r\sin \theta d\phi.$$

\noindent
Then, denoting the Minkowski components of the energy tensor by a bar,
we have

$$\bar T^0_0=T^0_0\,;\qquad\,
\bar T^1_1=T^1_1\,;\qquad\,\bar T^2_2=T^2_2\,;\qquad\,
\bar T^3_3=T^3_3.$$

\noindent
Next, we suppose that when viewed by a comoving with the fluid observer, the physical
content  of space consists of an anisotropic fluid of energy density $\mu$,
radial pressure $P_r$, and tangential pressure $P_\bot$.  Thus,  the covariant energy--momentum tensor in (local)
Minkowski coordinates is

\[ \left(\begin{array}{cccc}
\mu     &  0  &   0     &   0    \\
0 &  P_r     &   0     &   0    \\
0       &   0       & P_\bot  &   0    \\
0       &   0       &   0     &   P_\bot
\end{array} \right). \]

\noindent
Then 
\begin{equation}
T^0_0= \bar T^0_0 =\mu,
\label{8'}
\end{equation}
\begin{equation}
T^1_1=\bar T^1_1 =-P_r,
\label{9}
\end{equation}
\begin{equation}
T^2_2=T^3_3=\bar T^2_2 =\bar T^3_3 =-P_{\bot},
\label{10}
\end{equation}

\noindent
and 
the field equations read:

\begin{equation}
\mu =-\frac{1}{8\pi}\left[-\frac{1}{r^2}+e^{-\lambda}
\left(\frac{1}{r^2}-\frac{\lambda'}{r} \right)\right],
\label{fieq00}
\end{equation}

\begin{equation}
P_r =-\frac{1}{8\pi}\left[\frac{1}{r^2} - e^{-\lambda}
\left(\frac{1}{r^2}+\frac{\nu'}{r}\right)\right],
\label{fieq11}
\end{equation}

\begin{eqnarray}
P_\bot = \frac{1}{32\pi}e^{-\lambda}
\left(2\nu''+\nu'^2 -
\lambda'\nu' + 2\frac{\nu' - \lambda'}{r}\right),
\label{fieq2233}
\end{eqnarray}

where  primes stand for  derivatives with respect
to $r$.
\noindent

From these last  expressions it a simple matter to find  the hydrostatic equilibrium  equation, which  reads
\begin{equation}
P'_r=-\frac{\nu'}{2}\left(\mu+P_r\right)+\frac{2\left(P_\bot-P_r\right)}{r}.\label{Prp}
\end{equation}

 This is the  generalized Tolman-Opphenheimer-Volkoff equation for anisotropic matter. 

Alternatively, using 
\begin{equation}
\nu' = 2 \frac{m + 4 \pi P_r r^3}{r \left(r - 2m\right)},
\label{nuprii}
\end{equation}
which follows from the field equations, we may write
\begin{equation}
P'_r=-\frac{(m + 4 \pi P_r r^3)}{r \left(r - 2m\right)}\left(\mu+P_r\right)+\frac{2\left(P_\bot-P_r\right)}{r},\label{ntov}
\end{equation}
where $m$ is the  mass function defined by:
\begin{equation}
R^3_{232}=1-e^{-\lambda}=\frac{2m}{r},
\label{rieman}
\end{equation}
or, equivalently as
\begin{equation}
m = 4\pi \int^{r}_{0} \tilde r^2 \mu d\tilde r.
\label{m}
\end{equation}
Next, the four--velocity  vector is  given by:
\begin{equation}
u^{\mu}=(e^{-\frac{\nu}{2}},0,0,0)\label{u},
\end{equation}

 from which we can calculate the four acceleration, $a^\alpha=u^\alpha_{;\beta}u^\beta$, whose ony non--vanishing component is:
\begin{equation}
 a_1=-\frac{\nu
^{\prime}}{2},
\label{15a}
\end{equation}

It will be convenient to write the energy--momentum tensor  (\ref{8'})-(\ref{10}) as:
\begin{equation}
T^{\mu}_{\nu}= \mu u^{\mu}u_{\nu}-  P
h^{\mu}_{\nu}+\Pi ^{\mu}_{\nu},
\label{24'}
\end{equation}
with 
\begin{eqnarray}
\Pi^{\mu}_{\nu}&=&\Pi(s^{\mu}s_{\nu}+\frac{1}{3}h^{\mu}_{\nu});\quad
P=\frac{\tilde P_{r}+2P_{\bot}}{3}\nonumber \\ &&\Pi=P_{r}-P_{\bot};\quad h^\mu_\nu=\delta^\mu_\nu-u^\mu u_\nu,
\label{varios}
\end{eqnarray}
and  $s^\mu$  being defined by
\begin{equation}
s^{\mu}=(0,e^{-\frac{\lambda}{2}},0,0)\label{ese},
\end{equation}

with the  properties
$s^{\mu}u_{\mu}=0$,
$s^{\mu}s_{\mu}=-1$.

\noindent
For the exterior of the fluid distribution, the spacetime is that of Schwarzschild:

\begin{equation}
ds^2= \left(1-\frac{2M}{r}\right) dt^2 - \frac{dr^2}{ \left(1-\frac{2M}{r}\right)} -
r^2 \left(d\theta^2 + \sin^2\theta d\phi^2 \right).
\label{Vaidya}
\end{equation}

\noindent
In order to match smoothly the two metrics above on the boundary surface
$r=r_\Sigma=constant$, we  require the continuity of the first and the second fundamental
forms across that surface, producing 
\begin{equation}
e^{\nu_\Sigma}=1-\frac{2M}{r_\Sigma},
\label{enusigma}
\end{equation}
\begin{equation}
e^{-\lambda_\Sigma}=1-\frac{2M}{r_\Sigma},
\label{elambdasigma}
\end{equation}
\begin{equation}
\left[P_r\right]_\Sigma=0,
\label{PQ}
\end{equation}
where, from now on, subscript $\Sigma$ indicates that the quantity is
evaluated on the boundary surface $\Sigma$.

Eqs. (\ref{enusigma}), (\ref{elambdasigma}), and (\ref{PQ}) are the necessary and
sufficient conditions for a smooth matching of the two metrics (\ref{metric})
and (\ref{Vaidya}) on $\Sigma$.

\subsection{The Riemann and the Weyl tensor}
 As is well known, the
Riemann tensor may be expressed through the Weyl tensor
$C^{\rho}_{\alpha
\beta
\mu}$, the  Ricci tensor $R_{\alpha\beta}$ and the scalar curvature $R$,
as:
$$
R^{\rho}_{\alpha \beta \mu}=C^\rho_{\alpha \beta \mu}+ \frac{1}{2}
R^\rho_{\beta}g_{\alpha \mu}-\frac{1}{2}R_{\alpha \beta}\delta
^\rho_{\mu}+\frac{1}{2}R_{\alpha \mu}\delta^\rho_\beta$$
\begin{equation}
-\frac{1}{2}R^\rho_\mu g_{\alpha
\beta}-\frac{1}{6}R(\delta^\rho_\beta g_{\alpha \mu}-g_{\alpha
\beta}\delta^\rho_\mu).
\label{34}
\end{equation}

In the spherically symmetric case, the magnetic part of the Weyl tensor vanishes and we can express the Weyl tensor in terms of its electric part ($E_{\alpha \beta}=C_{\alpha \gamma \beta
\delta}u^{\gamma}u^{\delta}$) as
\begin{equation}
C_{\mu \nu 8\pi \lambda}=(g_{\mu\nu \alpha \beta}g_{\kappa \lambda \gamma
\delta}-\eta_{\mu\nu \alpha \beta}\eta_{\kappa \lambda \gamma
\delta})u^\alpha u^\gamma E^{\beta \delta},
\label{40}
\end{equation}
with $g_{\mu\nu \alpha \beta}=g_{\mu \alpha}g_{\nu \beta}-g_{\mu
\beta}g_{\nu \alpha}$,   and $\eta_{\mu\nu \alpha \beta}$ denoting the Levi--Civita tensor.
Observe that $E_{\alpha \beta}$ may also be written as 
\begin{equation}
E_{\alpha \beta}=E (s_\alpha s_\beta+\frac{1}{3}h_{\alpha \beta}),
\label{52bisx}
\end{equation}
with
\begin{equation}
E=-\frac{e^{-\lambda}}{4}\left[ \nu ^{\prime \prime} + \frac{{\nu
^{\prime}}^2-\lambda ^{\prime} \nu ^{\prime}}{2} -  \frac{\nu
^{\prime}-\lambda ^{\prime}}{r}+\frac{2(1-e^{\lambda})}{r^2}\right],
\label{defE}
\end{equation}
satisfying the following properties:
 \begin{eqnarray}
 E^\alpha_{\,\,\alpha}=0,\quad E_{\alpha\gamma}=
 E_{(\alpha\gamma)},\quad E_{\alpha\gamma}u^\gamma=0.
  \label{propE}
 \end{eqnarray} 

\subsection{The mass function and the Tolman mass}
Here we shall introduce the two most commonly used definitions for the mass of a sphere interior to the surface $\Sigma$, as well as some interesting relationships between them and the Weyl tensor. These will be used later to justify our choice for the complexity factor.
\subsubsection{The mass function}

Using  (\ref{Efeq}), (\ref{34}), (\ref{52bisx})   and the definition of the mass function given in (\ref{rieman}) (or (\ref{m})), we may write
 
\begin{equation}
m = \frac{4\pi}{3} r^3 \left(\mu + P_\bot - P_r\right) + \frac{r^3 E}{3},
\label{mW}
\end{equation}

from which  it is easy to obtain

\begin{equation}
 E= - \frac{4\pi}{r^3} \int^r_0{\tilde r^3 \mu' d\tilde r} + 
4\pi \left(P_r - P_\bot\right).
\label{Wint}
\end{equation}

\noindent
Finally, inserting (\ref{Wint}) into (\ref{mW}) we obtain

\begin{equation}
m(r) = \frac{4\pi}{3} r^3 \mu - 
\frac{4\pi}{3} \int^r_0{\tilde r^3 \mu 'd\tilde r}.
\label{mT00}
\end{equation}

Equation  (\ref{Wint}) relates the Weyl tensor to two fundamental physical properties of the fluid distribution, namely: density inhomogeneity and local anisotropy of pressure, whereas  (\ref{mT00}) expresses the mass function in  terms of its value in the case of a homogeneous energy density distribution, plus the change induced by density inhomogeneity.
\subsubsection{Tolman mass}
An alternative definition to  describe the energy content  of a  fluid sphere  was proposed by Tolman many years ago.
\noindent
The Tolman mass for a spherically symmetric static distribution 
of matter is given by  \cite{11}

\begin{eqnarray}
m_T =    4\pi \int^{r_\Sigma}_{0}{r^2 e^{(\nu+\lambda)/2} 
\left(T^0_0 - T^1_1 - 2 T^2_2\right) dr}.
\label{Tol}
\end{eqnarray}

\noindent
 Although Tolman's formula was introduced 
as a measure of the total energy of the system, with no commitment 
to its localization, we shall define the mass within a sphere of 
radius $r$, completely inside $\Sigma$, as 

\begin{eqnarray}
m_T =   4\pi \int^{r}_{0}{\tilde r^2 e^{(\nu+\lambda)/2} 
\left(T^0_0 - T^1_1 - 2 T^2_2\right) d\tilde r}\nonumber .
\label{Tolin}
\end{eqnarray}

\noindent
This extension of the global concept of energy to a local level 
 is suggested by the conspicuous role played by 
$m_T$ as the ``active gravitational mass'', which will be 
exhibited below.

In fact, it can be shown after some lengthy calculations (see \cite{13, 14} for details), that 
 
\begin{equation}
m_T  = e^{(\nu + \lambda)/2} 
\left[m(r) + 4\pi r^3 P_r\right].
\label{I+II}
\end{equation}

\noindent
Or, using the field equations

\begin{equation}
m_T = e^{(\nu - \lambda)/2} \, \nu' \, \frac{r^2}{2}.
\label{mT}
\end{equation}

\noindent
This last equation brings out the physical meaning of $m_T$ as the 
active gravitational mass. Indeed, as it follows from (\ref{15a}), the gravitational acceleration ($a=-s^\nu a_\nu$) of a test particle, 
instantaneously at rest in a static gravitational field,  is given by 

\begin{equation}
a = \frac{e^{- \lambda/2} \, \nu'}{2} =  \frac{e^{-\nu/2}m_T}{r^2} .
\label{a}
\end{equation}

Another expression for $m_T$, 
which appears to be more suitable for the discussion in Sec.IV is  (see \cite{13, 14} for details, but notice  slight changes in notation): 

\begin{widetext}
\begin{eqnarray}
m_T  =  (m_T)_\Sigma \left(\frac{r}{r_\Sigma}\right)^3 
 -  r^3 \int^{r_\Sigma}_r{e^{(\nu+\lambda)/2} \left[\frac{8\pi}{\bar r} 
\left(P_\bot - P_r \right)
+ \frac{1}{\bar r^4} \int^{\bar r}_0 {4\pi \tilde{r}^3 \mu' d\tilde{r}} 
 \right] d\bar r} ,
\label{emtebisbis}
\end{eqnarray}
\end{widetext}
or, using (\ref{Wint})

\begin{widetext}
\begin{eqnarray}
m_T  =  (m_T)_\Sigma \left(\frac{r}{r_\Sigma}\right)^3 
 -  r^3 \int^{r_\Sigma}_r{\frac{e^{(\nu+\lambda)/2}}{\tilde r} \left[4\pi
\left(P_\bot - P_r \right)
-E\right]d\tilde r}.
\label{emtebisbisb}
\end{eqnarray}
\end{widetext}
The important point to stress here is that the second  integral in  (\ref{emtebisbis}) describes the contribution of density inhomogeneity and local anisotropy of pressure to the Tolman mass.

We shall next present the orthogonal splitting of the Riemann tensor, and express it in terms of the variables considered so far.

\section{THE ORTHOGONAL SPLITTING OF THE RIEMANN TENSOR}
The orthogonal splitting of the Riemann tensor was first considered by Bel \cite{18}, here we shall follow closely  (with some changes) the notation  in  \cite{19}.

Thus following Bel, let us introduce the following tensors:
\begin{equation}
Y_{\alpha \beta}=R_{\alpha \gamma \beta \delta}u^{\gamma}u^{\delta},
\label{electric}
\end{equation}
\begin{equation}
Z_{\alpha \beta}=^{*}R_{\alpha \gamma \beta
\delta}u^{\gamma}u^{\delta}= \frac{1}{2}\eta_{\alpha \gamma
\epsilon \mu} R^{\epsilon \mu}_{\quad \beta \delta} u^{\gamma}
u^{\delta}, \label{magnetic}
\end{equation}
\begin{equation}
X_{\alpha \beta}=^{*}R^{*}_{\alpha \gamma \beta \delta}u^{\gamma}u^{\delta}=
\frac{1}{2}\eta_{\alpha \gamma}^{\quad \epsilon \mu} R^{*}_{\epsilon
\mu \beta \delta} u^{\gamma}
u^{\delta},
\label{magneticbis}
\end{equation}
where $*$ denotes the dual tensor, i.e.
$R^{*}_{\alpha \beta \gamma \delta}=\frac{1}{2}\eta_{\epsilon \mu \gamma \delta}R_{\alpha \beta}^{\quad \epsilon \mu}$.

It can be shown that the Riemann tensor  can be expressed through  these tensors in what is called the orthogonal splitting of the Riemann tensor (see \cite{19} for details).
However, instead of using the explicit form of the splitting of Riemann tensor (eq.(4.6) in \cite{19}), we shall proceed as follows (for details see \cite{20}, where the general non--static case has been considered). 

Using the Einstein equations we may write (\ref{34}) as 
\begin{equation}
R^{\alpha \gamma}_{\quad \beta\delta}=C^{\alpha\gamma}_{\quad
\beta \delta}+28\pi T^{[\alpha}_{\,\,
[\beta}\delta^{\gamma]}_{\,\, \delta]}+8\pi T(\frac{1}{3} \delta
^{\alpha}_{\,\, [\beta}\delta^{\gamma}_{\,\, \delta]}-\delta
^{[\alpha}_{\quad [\beta}\delta^{\gamma]}_{\,\,
\delta]}),\label{RiemannT}
\end{equation}
then feeding back (\ref{24'}) into  (\ref{RiemannT}) we split the Riemann tensor as 
\begin{equation}
R^{\alpha \gamma}_{\,\, \beta \delta}=R^{\alpha
\gamma}_{(I)\,\,\beta\delta}+R^{\alpha
\gamma}_{(II)\,\,\beta\delta}+R^{\alpha\gamma}_{(III)\,\,\beta\delta},
\label{Riemann}
\end{equation}
where
\begin{eqnarray}
 R^{\alpha \gamma}_{(I)\,\,\beta \delta}=16\pi  \mu
u^{[\alpha}u_{[\beta}\delta^{\gamma]}_{\,\,\delta]}-28\pi 
Ph^{[\alpha}_{\,\,[\beta}\delta^{\gamma]}_{\,\, \delta]}\nonumber \\+8\pi
( \mu-3
P)(\frac{1}{3}\delta^{\alpha}_{\,\,[\beta}\delta^{\gamma}_{\,\,\delta]}
-\delta^{[\alpha}_{\,\,[\beta}\delta^{\gamma]}_{\,\,\delta]})
\end{eqnarray}
\begin{eqnarray}
R^{\alpha \gamma}_{(II)\,\,\beta\delta}=16\pi (\Pi^{[\alpha}_{\,\,
[\beta}\delta^{\gamma]}_{\,\, \delta]}
\end{eqnarray}
\begin{equation}
R^{\alpha
\gamma}_{(III)\,\,\beta\delta}=4u^{[\alpha}u_{[\beta}E^{\gamma]}_{\,\,\,
\delta]}-\epsilon^{\alpha \gamma}_{\quad
\mu}\epsilon_{\beta\delta\nu}E^{\mu\nu}
\label{RiemannH}
\end{equation}
with
\begin{eqnarray}
\epsilon_{\alpha\gamma\beta}=u^\mu\eta_{\mu\alpha\gamma\beta},\quad
\epsilon_{\alpha\gamma\beta}u^\beta=0 \label{conn},
 \end{eqnarray} 
and where the vanishing, due to the spherical symmetry, of the  magnetic part of the Weyl tensor ($H_{\alpha \beta}=^{*}C_{\alpha \gamma \beta
\delta}u^{\gamma}u^{\delta}$)  has been used.

Using the results above, we can now  find the explicit expressions for the three tensors $Y_{\alpha \beta}, Z_{\alpha \beta}$ and  $X_{\alpha \beta}$ in terms of  the physical variables, we obtain

\begin{equation}
Y_{\alpha\beta}=\frac{4\pi}{3}(\mu+3
P)h_{\alpha\beta}+4\pi \Pi_{\alpha\beta}+E_{\alpha\beta},\label{Y}
\end{equation}

\begin{equation}
Z_{\alpha\beta}=0,\label{Z}
\end{equation}

and

\begin{equation}
X_{\alpha\beta}=\frac{8\pi}{3} \mu
h_{\alpha\beta}+4\pi
 \Pi_{\alpha\beta}-E_{\alpha\beta}.\label{X}
\end{equation}

As shown in \cite{20}, the tensors  above may be expressed in terms of some scalar functions, referred to as structure scalars.

Indeed,    from the  tensors  $X_{\alpha \beta}$ and $Y_{\alpha \beta}$   we may define four scalars functions, in terms of which these tensors may be written, these are denoted by $X_T, X_{TF},  Y_T, Y_{TF}$, a fifth scalar associated to the tensor $Z_{\alpha \beta}$ vanishes in the static case (see \cite{20} for details). 

These scalars may written as:
\begin{equation}
 X_T=8\pi  \mu,
\label{esnIII}
\end{equation}
  
\begin{equation}
X_{TF}= 4\pi \Pi-E,
\label{defXTF}
\end{equation}
or using (\ref{Wint})
\begin{equation}
X_{TF}= \frac{4\pi}{r^3} \int^r_0{\tilde r^3 \mu' d\tilde r},
\label{defXTFbis}
\end{equation}

\begin{equation}
Y_T=4\pi( \mu+3 P_r-2\Pi),
\label{esnV}
\end{equation}

\begin{equation}
Y_{TF}= 4\pi \Pi+E,
\label{defYTF}
\end{equation}
or using (\ref{Wint})
\begin{equation}
Y_{TF}=8\pi \Pi- \frac{4\pi}{r^3} \int^r_0{\tilde r^3 \mu' d\tilde r}.
\label{defYTFbis}
\end{equation}

From the above it follows that local anisotropy of pressure is  determined by  $X_{TF}$ and $Y_{TF}$
by
\begin{equation}
8\pi \Pi=X_{TF} + Y_{TF}.
\label{defanisxy}
\end{equation}

To establish the physical meaning of  $Y_T$  and $Y_{TF}$ let us get back to equations (\ref{emtebisbis}) or (\ref{emtebisbisb}), using (\ref{defYTF}) or (\ref{defYTFbis}) we get

\begin{eqnarray}
m_T  =  (m_T)_\Sigma \left(\frac{r}{r_\Sigma}\right)^3
 +  r^3 \int^{r_\Sigma}_r{\frac{e^{(\nu+\lambda)/2}}{\tilde r} Y_{TF}d\tilde r} .
\label{emtebis}
\end{eqnarray}

 Comparing the above expression with (\ref{emtebisbis}) we see that $Y_{TF}$ describes the influence of the local anisotropy of pressure and density inhomogeneity on the Tolman mass. Or, in other words, $Y_{TF}$ describes how these two factors modify the value of the Tolman mass, with respect to its value for the homogeneous isotropic fluid. It is also worth recalling that $Y_{TF}$,  together with $X_{TF}$, determines the local anisotropy of the fluid distribution.

Finally, observe that the Tolman mass 
 may be written as
\begin{eqnarray}
m_T = & &   \int^{r}_{0}{\tilde r^2 e^{(\nu+\lambda)/2}
Y_Td\tilde r}.
\label{TolinIII}
\end{eqnarray}
\section{The complexity factor}
We are now ready to introduce our definition of complexity, which as mentioned before, will be represented by a single scalar function referred to as the complexity factor.

Following the tradition established  in previous works on the concept of complexity,  we shall start by defining the simplest (the less complex) system, within  the whole space of exact solutions of the Einstein equations for spherically symmetric static fluid distributions with anisotropic pressure. 

 For simple intuitive reasons, we shall assume that at least one of these simplest systems is represented by a homogeneous energy density distribution with isotropic pressure. For such a system, as we have seen in the previous section, the structure scalar $Y_{TF}$ vanishes. Furthermore, this single scalar function, encompasses all the modifications produced  by the energy density inhomogeneity and the anisotropy of the pressure, on the active  gravitational (Tolman) mass. 

From the comments above, it appears well justified to identify the complexity factor with $Y_{TF}$. 

The following remarks are in order at this point:
\begin{itemize}
\item It should be noticed that the complexity factor so defined, not only vanishes for the homogeneous, isotropic fluid, where the two terms in (\ref{defYTFbis}) vanish identically, but also for all configurations where the two terms in (\ref{defYTFbis}) cancel each other.
\item From the point above it follows that there are a wealth of configurations satisfying the vanishing  complexity conditions.
\item It is worth noticing that whereas the contribution of the pressure anisotropy to $Y_{TF}$ is local, the contribution of the density energy inhomogeneity is not.
\item If we allow the fluid distribution to be electrically charged, then the corresponding $Y_{TF}$ will include contributions from the electric charge which are of, both,  local and non--local nature (see eq.(25) in \cite{21}).
\end{itemize}

In the next section, we shall present two  examples of inhomogeneous and anisotropic fluid configurations, satisfying the vanishing complexity factor condition.

\section{Fluid distributions with vanishing complexity factor}
The Einstein equations for  a spherically symmetric static, anisotropic fluid (Eqs.(\ref{fieq00})--(\ref{fieq2233}), form  a system of three ordinary differential equations for five unknow functions ($\nu, \lambda, \mu, P_r, P_\bot$). Accordingly, if we impose the condition $Y_{TF}=0$ we shall need still one condition in order to solve the system. 

The vanishing complexity factor condition, according to (\ref{defYTFbis})) reads:

\begin{equation}
 \Pi=\frac{1}{2r^3} \int^r_0{\tilde r^3 \mu' d\tilde r}.
\label{vcfc}
\end{equation}

It is worth noticing that, as it follows from (\ref{vcfc}), the vanishing complexity factor condition implies either, homogeneous energy density and pressure isotropy, or inhomogeneous energy density and pressure anisotropy.

Also, it should be noticed that (\ref{vcfc} may be regarded as a non--local equation of state, somehow similar to the one proposed some years ago in \cite{21bis}.

Just for the sake of illustration, here we shall propose some examples.

\subsection{The Gokhroo and Mehra ansatz}
A family of anisotropic spheres has been found in \cite{22}, which lead to physically satisfactory models for compact objects.

The starting point  for the obtention of these models, is an assumption on the form of the metric function $\lambda$ which reads

\begin{equation}
e^{-\lambda}=1-\alpha r^2+\frac{3K\alpha r^4}{5r_\Sigma^2},
\label{lgm}
\end{equation}
producing, because of (\ref{fieq00}) and (\ref{m})
\begin{equation}
\mu=\mu_0\left(1-\frac{Kr^2}{r_\Sigma^2}\right)
\label{mugm}
\end{equation}
and
\begin{equation}
m(r)=\frac{4\pi\mu_0r^3}{3}\left(1-\frac{3Kr^2}{5r_\Sigma^2}\right),
\label{mgm}
\end{equation}

where $K$ is a constant in the range $(0, 1)$ and $\alpha=\frac{8\pi \mu_0}{3}$.

Next, from (\ref{fieq11}), (\ref{fieq2233}), we may write
\begin{eqnarray}
8\pi(P_r-P_\bot)&=&e^{-\lambda}\left[-\frac{\nu^{\prime \prime}}{2}-(\frac{\nu^\prime}{2})^2+\frac{\nu^\prime}{2r}+\frac{1}{r^2}+\frac{\lambda^\prime}{2}(\frac{\nu^\prime}{2} +\frac{1}{r})\right]\nonumber \\&-&\frac{1}{r^2}
 \label{anis}
\end{eqnarray}

\noindent Then, introducing the variables 
\begin{equation}
e^{\nu (r)}=e^{\int (2z(r)-2/r)dr},
\label{v1}
\end{equation}

and 
\begin{equation}
e^{-\lambda}=y(r),
\label{v2}
\end{equation}
and feeding back into  (\ref{anis}) we get:

\begin{equation}
y^{\prime}+y\left[\frac{2z^\prime}{z}+2z-\frac{6}{r}+\frac{4}{r^2
z}\right]=-\frac{2}{z}\left(\frac{1}{r^2}+\frac{\Pi(r)}{8\pi}\right). \label{eq1}
\end{equation}

In our case, $y(r)$ is given by (\ref{lgm}) whereas $\Pi$ is obtained from (\ref{vcfc}) and (\ref{mugm}). Feeding back these expressions in (\ref{eq1}), this last equation becomes a Ricatti equation, whose integration provides $z$, and thereby  the solution is completely determined.

Indeed, in terms of these two functions $z, \Pi$, the line element becomes (see \cite{16} for details):

\begin{widetext}
\begin{equation}
ds^2=-e^{\int (2z(r)-2/r)dr}dt^2+\frac{z^2(r) e^{\int(\frac{4}{r^2
z(r)}+2z(r))dr}} {r^6(-2\int\frac{z(r)(1+\frac{\Pi (r)r^2}{8\pi})
e^{\int(\frac{4}{r^2
z(r)}+2z(r))dr}}{r^8}dr+C)}dr^2+r^2d\theta^2+r^2sin^2\theta
d\phi^2. \label{metric2}
\end{equation}
\end{widetext}
where $C$ is a constant of integration.

And for  the physical variables  we have:

\begin{equation}
4\pi P_r=\frac{z(r-2m)+m/r-1}{r^2}\label{Pr},
\end{equation}
\noindent 

 \begin{equation}
4\pi \mu =\frac{m^{\prime}}{r^2}\label{rho},
\end{equation}
and 
\begin{equation}
4\pi P_\bot=(1-\frac{2m}{r})(z^{\prime}+z^2-\frac{z}{r}+\frac{1}{r^2})+z(\frac{m}{r^2}-\frac{m^{\prime}}{r}).
\label{Pbot}
\end{equation}

The so obtained solution is regular at the origin, and satisfies the conditions $\mu>0$, $\mu>P_r, P_\bot$.

Also, to avoid singular behaviour of physical variables on the boundary of the source ($\Sigma$), the solution should also satisfy the Darmois conditions on the boundary  (\ref{enusigma}), (\ref{elambdasigma}), (\ref{PQ}). 
\subsection{The polytrope with vanishing complexity factor}
The polytropic equation of state plays an important role in the study of self--gravitating systems, both, in Newtonian and general relativistic astrophysics. The study of polytropes for anisotropic matter has been considered in detail in the recent past \cite{23, 24, 25}. 

After, adopting the polytropic equation of state, in the case of anisotropic matter, we still need an additional condition in order to solve the corresponding system of equations.

 Here we propose to complement the polytropic equation of state with the vanishing   complexity factor condition.

Thus our model is obtained on the basis of the following conditions:
\begin{equation}P_r=K\mu^{\gamma}=K\mu^{1+1/n};\qquad Y_{TF}=0,
\label{p2}\end{equation} 
where constants $K$, $\gamma$, and $n$ are usually called  the polytropic constant, polytropic exponent, and polytropic index, respectively.

From the polytropic equation of state we obtain two equations which read:

\begin{widetext}

\begin{eqnarray}
\xi^2 \frac{d\Psi}{d\xi}\left[\frac{1-2(n+1)\alpha v/\xi}{1+\alpha \Psi}\right]+v+\alpha\xi^3 \Psi^{n+1}+\frac{2\Pi \Psi^{-n}\xi}{P_{rc}(n+1)} \left[\frac{1-2\alpha(n+1)v/\xi}{1+\alpha \Psi}\right]=0,\label{TOV2anis_WB}
\end{eqnarray}
\end{widetext}
and 
\begin{equation}\frac{dv}{d\xi}=\xi^2 \Psi^n,\label{veprima2}\end{equation}
where
\begin{equation}
\alpha=P_{rc}/\mu_{c},\quad r=\xi/A,  \quad A^2=4 \pi \mu_{c}/\alpha (n+1)\label{alfa},\end{equation}

\begin{equation}\Psi^n=\mu/\mu_{c},\quad v(\xi)=m(r) A^3/(4 \pi\mu_{c}),\label{psi}\end{equation}
where subscript $c$ indicates that the quantity is evaluated at the center. At the boundary surface $r=r_\Sigma$ ($\xi=\xi_\Sigma$) we have $\Psi(\xi_\Sigma)=0$ (see \cite{25} for details).

Equations (\ref{TOV2anis_WB}), (\ref{veprima2}), form a system of two first order  ordinary differential equations for the three unknown functions: $\Psi, v, \Pi$, depending on a duplet of parameters $n, \alpha$. In order to proceed further with the modeling of a compact object,  we shall further assume the vanishing complexity factor condition, which with the notation above, reads:
\begin{equation}
\frac{6 \Pi}{n\mu_c}+\frac{2\xi}{n\mu_c}\frac{d\Pi}{d\xi}=\Psi^{n-1}\xi \frac{d\Psi}{d\xi}.
\label{pol3}
\end{equation}

Now we have a system of three ordinary differential equations (\ref{TOV2anis_WB}), (\ref{veprima2}), (\ref{pol3}) for the three unknown functions $\Psi, v, \Pi$, which may be integrated for an arbitrary duplet of values of the parameters $n, \alpha$, only constrained by the physical conditions (see \cite{24} for details):
\begin{equation}
\mu>0, \qquad \alpha \Psi \leq 1, \qquad \frac{3v}{\xi^3 \Psi^n}+\alpha \Psi-1 \leq1.
\label{conditionsIII}
\end{equation}

Since we do not intend to present specific models of compact object, we shall not proceed further to integrate the above equations.

Finally, it is worth mentioning that the generalization of the Newtonian polytrope to the  general relativistic case, admits two possibilities. One is the equation (\ref{p2}), the other is $P_r=K\mu_{b}^{\gamma}=K\mu_b^{1+1/n}$, where $\mu_{b}$ denotes the baryonic (rest) mass density. The treatment of this last case has been described in detail in \cite{24}. The equations equivalent to (\ref{TOV2anis_WB}) and( \ref{pol3}), in this case are:
\begin{widetext}

\begin{eqnarray}
\xi^2 \frac{d\Psi_b}{d\xi}\left[\frac{1-2(n+1)\alpha v/\xi}{1+\alpha \Psi_b}\right]+v+\alpha\xi^3 \Psi_b^{n+1}+\frac{2\Pi \Psi_b^{-n}\xi}{P_{rc}(n+1)} \left[\frac{1-2\alpha(n+1)v/\xi}{1+\alpha \Psi_b}\right]=0,\label{TOV1anis_WB}
\end{eqnarray}
\end{widetext}

\begin{equation}
\frac{6 \Pi}{n\mu_{bc}}+\frac{2\xi}{n\mu_{bc}}\frac{d\Pi}{d\xi}=\Psi_b^{n-1}\xi \frac{d\Psi_b}{d\xi}\left[1+K(n+1)\mu_{bc}^{1/n} \Psi_b\right],
\label{pol4}
\end{equation}
with $\Psi_b^n=\mu_b/\mu_{bc}$.

\section{conclusions}
We have introduced a new concept of complexity, for static spherically symmetric relativistic fluid distributions, which stems from the basic assumption that one of the less complex systems corresponds to an homogeneous (in the energy density) fluid distribution with isotropic pressure. Then, as an obvious candidate to measure the degree of complexity, appears the structure scalar $Y_{TF}$. We would like to emphasize here, the reasons behind such a proposal:
\begin{itemize}
\item The scalar function $Y_{TF}$  contains contributions from the energy density inhomogeneity and the local pressure anisotropy, combined in a very specific way.
\item This scalar measures the departure of the value of the Tolman mass for the homogeneous and anisotropic fluid, produced by the energy density inhomogeneity and the pressure anisotropy.
\item In the case of a charged fluid,  this scalar also encompasses the effect of the electric charge.
\item In the general non--static, dissipative fluid distribution, $Y_{TF}$ contains, besides the contributions from the energy density inhomogeneity and the local pressure anisotropy, also contributions from the dissipative fluxes. 
\item In the non--static case,  the vanishing of $Y_{TF}$ is a necessary condition for the stability of the shear--free condition \cite{26}. This last condition  generalizes to the relativistic case the homologous evolution, which in turn appears to be the more ``orderly'' type of evolution (see \cite{27} for a discussion on this point).
\end{itemize}

Next, we have exhibited some exact solutions satisfying the vanishing complexity factor condition. As mentioned before, the intention was not to provide models with specific astrophysical interest, but just illustrate how such models may be found, with two examples. 

Among the many existing possible conditions  to complement the vanishing complexity factor condition, in order to obtain models, we may mention:
\begin{itemize}
\item $P_r=0.$
\item $P_{\bot}=0$.
\item The non--local equation of state proposed in \cite{21bis}.
\item The ansazts proposed by Cosenza et al. in \cite{28}.
\item The Karmarkar condition \cite{29}.
\item The Krori--Barua ansatz \cite{30}.
\end{itemize}
Before ending,   we would like to make some final remarks and to present a partial list of issues, which  remain  unanswered in this manuscript, but  should be addressed in the future. 
\begin{itemize}
\item The definition of complexity proposed in this work, is not directly related to entropy or disequilibrium, although  it is possible that such a link might exist after all. If so, how  could, such relationship,  be brought out?
\item We have introduced  a definition for complexity, but we have not explored how such a concept is related to physical relevant properties of the source such as the stability, or the maximal degree of compactness. 
\item The complexity factor may be negative. If we assume that the simplest systems are described by a vanishing complexity factor, what is the physical meaning of a negative complexity factor?
\item How does the complexity factor evolves? Do physically meaningful systems prefer vanishing complexity factor?
\item Since, $Y_{TF}$ has been defined also for the general non--static case, and for the non--spherical case, the extension of the discussion started here, to these cases, may proceed without much difficulty. However it should  be mentioned that in this latter case (non--spherical) the number of structure scalars is larger, and it is possible that complexity should be defined through more than one scalar function.
\item The complexity factor for a charged (spherically symmetric) fluid is known, but what is the complexity factor for a different type of field (e.g. a  scalar field?).
\item We have defined the complexity factor in the context of the Einstein theory of gravitation, but how should we define it, in the context of any other alternative theory?
\end{itemize}
\
\begin{acknowledgments}

This  work  was partially supported by the Spanish  Ministerio de Ciencia e
Innovaci\'on under Research Projects No.  FIS2015-65140-P (MINECO/FEDER).

\end{acknowledgments}


\begin{thebibliography}{88}
\bibitem{0}A. N. Kolmogorov, {\it Prob. Inform. Theory J.} {\bf 1}, 3 (1965).
\bibitem{1} P. Grassberger, {\it Int. J. Theor. Phys.} {\bf  25}, 907 (1986).
\bibitem{1bis}S. Lloyd and H. Pagels, {\it Ann. Phys.} {\bf 188}, 186 (1988).
\bibitem{2} J. P. Crutchfield and Karl Young, {\it Phys. Rev. Lett.} {\bf 63}, 105 (1989).
\bibitem{2bis}P. W. Anderson, {\it Physics Today} {\bf 7}, (1991).
\bibitem{2bisbis} G. Parisi, {\it Phys. World} {\bf 6}, 42 (1993).
\bibitem{3}R. Lopez--Ruiz, H. L. Mancini, and X. Calbet, {\it Phys. Lett. A} {\bf 209}, 321 (1995).
\bibitem{3bis} D. P. Feldman and J. P. Crutchfield,  {\it Phys. Lett. A} {\bf 238}, 244 (1998).
\bibitem{3bisbis} X. Calbet and R. Lopez--Ruiz, {\it Phys. Rev. E} {\bf 63}, 066116 (2001).
\bibitem{4}R. G. Catalan, J. Garay and R. Lopez--Ruiz, {\it Phys. Rev. E} {\bf 66}, 011102 (2002).
\bibitem{4bis} J. Sa\~nudo and R. Lopez--Ruiz, {\it Phys. Lett. A} {\bf 372}, 5283 (2008).
\bibitem{6bis}C. P. Panos, N. S. Nikolaidis, K. Ch. Chatzisavvasand and C. C. Tsouros {\it Phys. Lett. A} {\bf 373}, 2343 (2009).
\bibitem{5}J. Sa\~nudo and A. F. Pacheco, {\it Phys. Lett. A} {\bf 373}, 807 (2009).
\bibitem{6}K. Ch. Chatzisavvas, V. P. Psonis, C. P. Panos and Ch. C. Moustakidis, {\it Phys. Lett. A} {\bf 373}, 3901 (2009).
\bibitem{7}M. G. B. de Avellar and J. E. Horvath, {\it Phys. Lett. A} {\bf 376}, 1085 (2012).
\bibitem{8}R. A. de Souza, M. G. B. de Avellar and J. E. Horvath, {\it arxiv: 1308.3519}.
\bibitem{9} M. G. B. de Avellar and J. E. Horvath, {\it arxiv: 1308.1033}.
\bibitem{10}M. G. B. de Avellar , R. A. de Souza, J. E. Horvath and D. M. Paret, {\it Phys. Lett. A} {\bf 378}, 3481 (2014).
\bibitem{11}R. Tolman, {\it  Phys. Rev.}  {\bf 35}, 875 (1930).
\bibitem{13} L. Herrera and N. O. Santos, {\it Phys. Rep.} {\bf 286}, 53
(1997).
\bibitem{14} L. Herrera, A. Di Prisco, J. Hern\'andez-Pastora, and N. O. Santos,
{\it Phys. Lett. A}  {\bf 237}, 113 (1998).
\bibitem{16} L. Herrera, J. Ospino and A. Di Prisco, {\it Phys.Rev. D} {\bf 77}, 027502 (2008).
\bibitem{17} H. Bondi, {\it Proc. R. Soc. London} {\bf A281}, 39 (1964).
\bibitem{18} L. Bel, {\it Ann. Inst. H
Poincar\'e}  {\bf 17}, 37 (1961).
\bibitem{19} A. Garc\'ia--Parrado Gomez Lobo, {\it arXiv:0707.1475v2}.
\bibitem{20}L. Herrera, J. Ospino, A. Di Prisco, E. Fuenmayor and  O. Troconis, {\it Phys. Rev. D} {\bf 79}, 064025 (2009).
\bibitem{21}L. Herrera, A. Di Prisco and J. Iba\~nez,  {\it Phys. Rev. D} {\bf 84}, 107501 (2011)
\bibitem{21bis} H. Hern\'andez, L. A.  N\'u\~nez, {\it Can. J. Phys.} {\bf 82}, 29 (2004).
\bibitem{22}  M. K. Gokhoo and A. L. Mehra,  {\it Gen. Relativ. Gravit.} {\bf 26}, 75 (1994).
\bibitem{23} L. Herrera and W. Barreto, {\it Phys. Rev. D} {\bf 87}, 087303, (2013).
\bibitem{24} L. Herrera and W. Barreto, {\it Phys. Rev. D} {\bf 88}, 084022, (2013).
\bibitem{25} L. Herrera, E. Fuemayor and P. Leon, {\it Phys. Rev. D} {\bf 93}, 024047, (2016).
 \bibitem{26}L. Herrera, A. Di Prisco and  J. Ospino, {\it  Gen. Relativ.  Gravit.}  {\bf 42}, 1585 (2010).
\bibitem{27}L. Herrera and   N. O. Santos, {\it Month. Not.  Roy. Astr. Soc.}, {\bf 343}, 1207 (2003).
\bibitem{28} M. Cosenza, L. Herrera, M. Esculpi and L. Witten, J. Math. Phys., {\bf 22}, 118 (1981).
\bibitem{29} K. R. Karmarkar, {\it Proc. Ind. Acad. Sci. A}  {\bf 27},  56 (1948).
\bibitem{30} K. D. Krori and J. Barua, {\it J. Phys. A: Math. Gen.} {\bf 8}, 508
(1975).
\end{thebibliography}
\end{document}